\documentclass[a4paper,11pt]{article}
\pdfoutput=1 

\usepackage{jheppub} 

\usepackage[T1]{fontenc} 
\usepackage{amsmath}
\usepackage{mathrsfs}

\title{\boldmath Holographic non-computers}

\author[a]{Jos\'e L.F. Barb\'on}
\author[a]{and Javier Mart\'in-Garc\'ia}

\affiliation[a]{Instituto de F\'isica Te\'orica IFT UAM/CSIC \\ Calle Nicol\'as Cabrera 13. UAM, Cantoblanco. 28049 Madrid, Spain}

\emailAdd{jose.barbon@uam.es}
\emailAdd{javier.martingarcia1@gmail.com}

\abstract{We introduce the notion of holographic non-computer as a system which exhibits parametrically large delays in the growth of complexity, as calculated within the Complexity-Action proposal. Some known examples of this behavior include extremal black holes and near-extremal hyperbolic black holes. Generic black holes  in higher-dimensional gravity also show non-computing features. Within the $1/d$ expansion of General Relativity, we show that large-$d$ scalings which  
capture the qualitative features of complexity, such as a linear growth regime and a plateau at exponentially long times, also exhibit
an initial computational delay proportional to $d$. While consistent for large AdS black holes, the required `non-computing' scalings are incompatible with thermodynamic stability for Schwarzschild black holes, unless they are tightly caged. }

\begin{document} 
\begin{flushright}
IFT-UAM/CSIC-17-101
\end{flushright}
\maketitle
\flushbottom
\setlength\parskip{5pt}
\setlength\abovedisplayskip{10pt}
\setlength\belowdisplayskip{10pt}
\setlength\abovedisplayshortskip{10pt}
\setlength\belowdisplayshortskip{10pt}

\newcommand{\MC}{\mathcal{C}}
\newcommand{\CC}{\mathcal{C}}
\newcommand{\CV}{\mathcal{V}}
\newcommand{\CB}{\mathcal{B}}

\section{Introduction}

Concepts and perspectives from the field of quantum information theory have been a fruitful  source ideas in the quest to gain a deeper understanding of holography, particularly in the context of AdS/CFT. One such  perspective is offered by the recent proposals of holographic quantum complexity. Roughly speaking, an extensive measure in the bulk geometry, such as extremal volume or on-shell action, is associated to the  complexity of the quantum state. 

Complexity itself can be loosely defined by the size of the minimum quantum circuit that is needed to approximate a state from a reference one as measured by some set of elementary entanglement operations. Expectations from quantum many-body theory predict a linear growth of such quantity for generic excited states. Motivated  by the study of Einstein-Rosen bridges, the initial conjecture implied a correspondence between the expected linear growth of quantum complexity for thermal systems and the growth of wormhole volumes (defined as extremal codimension-one surfaces) for AdS black holes \cite{SusskindEntnotEnough}. This conjecture became further developed into the so called Action-Complexity proposal, which postulates an equivalence between quantum complexity and the gravitational on-shell action on a particular region of space called the Wheeler-DeWitt patch \cite{BrownSusskindAction} (see also \cite{SusskindCompBHHor, StanfordShockWave, SusskindCAcorto, Susskind2law, Cesar}).

The linear growth of complexity, in either the Complexity-Volume or Complexity-Action prescription,
\begin{equation}
{dC_V \over dt} = c \,S\,T\,,\qquad {dC_A \over dt} = 2M\;,
\end{equation}
with $c$ a parameter of order unity, is an  asymptotic statement which should hold at times parametrically larger than the scrambling time, but still 
smaller than the Heisenberg time of the system, which scales exponentially with the entropy,  $t_{\rm H} \sim T^{-1} \exp(S) $.  At times of this order the complexity itself is expected to saturate, fluctuating around  a maximum value of order $\exp(S)$ (cf.  \cite{Susskindtypical}). 

In this paper we study various situations in which the picture of a linear complexity growth followed by a long-time saturation plateau
is  modified by a  `computational delay'.  In the examples we consider, this delay is either permanent or rather it can be made parametrically large in some approximation, representing an instance of a `holographic non-computer'. The non-computing behavior is characteristic of the Complexity-Action proposal (CA), which is perhaps more intimately related to the `computational' aspect of holographic complexity, as opposed to the Complexity-Volume proposal (CV), which is closer to the interpretation of complexity in terms of the tensor-network parametrization of quantum states.

In section 2 of this paper we consider the  simplest  example of  a holographic non-computer, namely the well-known case of extremal black holes, whose complexity remains constant for arbitrarily long times \cite{BrownSusskindAction}. This is however quite a peculiar example at zero temperature, where our general intuition of quantum complexity might not hold. In section 3 we consider a less fine-tuned example  provided by small hyperbolic black holes, which exhibit again a \textit{non-computer} behavior for the finite range of temperatures that span its near-extremal regime \cite{MyersFormation}.

Finally, we devote section 4 to a third class of non-computing behavior,  exhibited by higher dimensional black holes. As showed in \cite{BrownSusskindAction}, black holes in four or higher spacetime dimensions enjoy a period of constant Action-Complexity at early times, postponing the usual linear growth after some delay lapse which depends on physical properties of the black hole. As we will show, this behavior gets enhanced as the dimension grows and can lead to an eternal non-computer system at leading order in a large $d$ expansion.

The large-$d$ expansion of General Relativity has illuminated a number of classical dynamical regimes in various black-hole systems (cf. \cite{EmparanlargeD, Minwalla}). While its status at the quantum level is rather unclear, we find it interesting that a non-trivial statement can be made for such highly quantum properties as the computational complexity of black holes.

\section{A frozen non-computer. Extremal black holes}
\label{sectionfrozen}

As it was shown by \cite{BrownSusskindAction, Cai, Poisson}, Reissner-N\"ordstrom black holes exhibit a constant computation rate which depends on both mass and charge of the solution. In particular, an explicit computation yields for the rate of growth of the on-shell action:

\begin{equation}
\dfrac{dI}{dt} = 2(d-1)V_h \left(\dfrac{Q^2}{r_-^{d-2}}-\dfrac{Q^2}{r_+^{d-2}} \right),
\end{equation}
where $Q$ is the charge of the black hole, $r_\pm$ the location of outer and inner horizons and $V_h$ the volume of the event horizon. 
Here we quote the results for the rescaled quantity
\begin{equation}
I = 16\pi G\, C_A\;,
\end{equation}
where $C_A$ is the on-shell gravitational action evaluated on the WdW patch. 

\begin{figure}[h]
\begin{center}
\includegraphics[height=7cm]{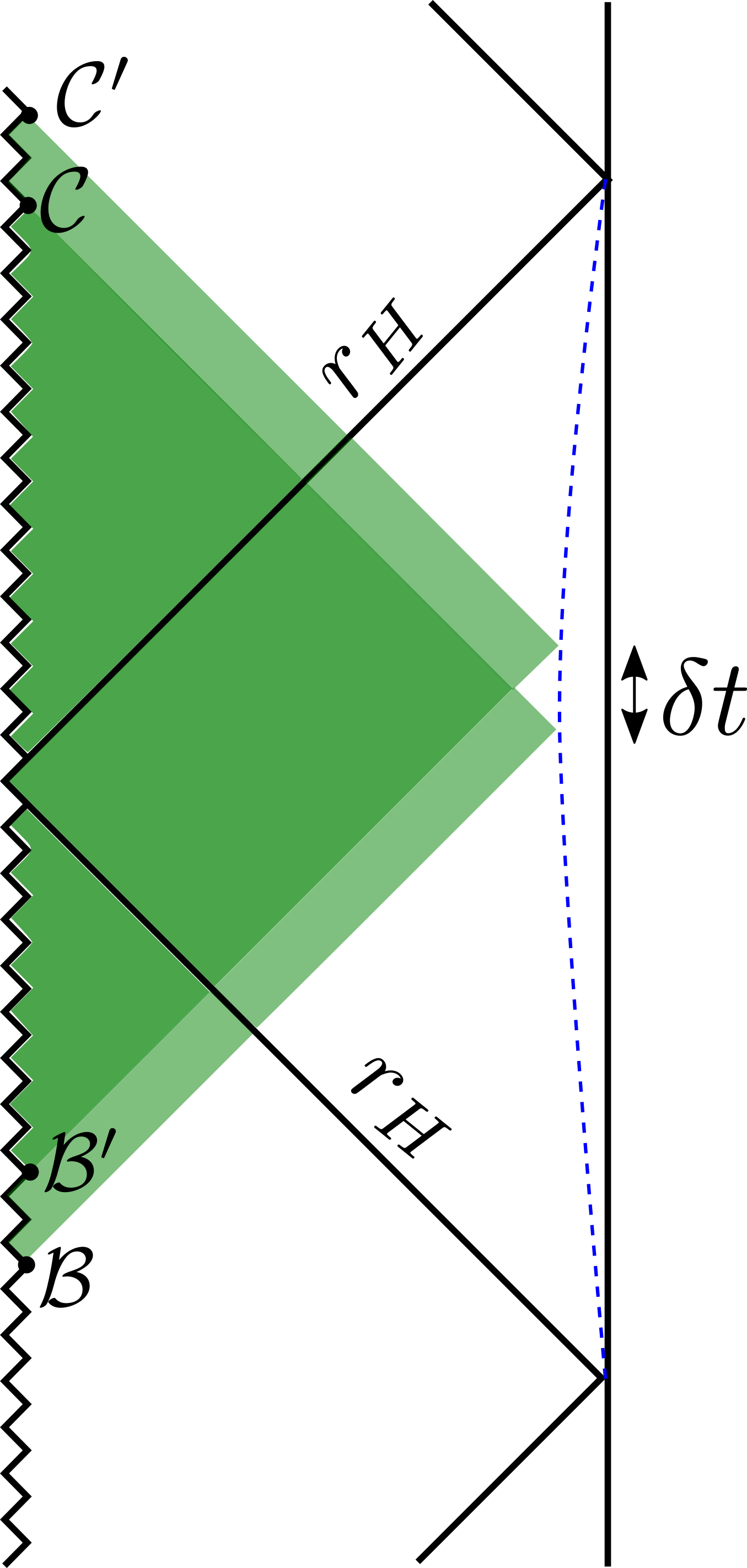}
\end{center}
\caption{WdW patch for a charged extremal black hole. The bulk volume and boundary contributions are conserved by the isometries of the spacetime, whereas the joint piece vanishes due to the sphere shrinking at the singurality. }
\label{Extremal}
\end{figure}

As we go to the extremal limit (taking $r_- \rightarrow r_+$ with constant $Q$) the quantity above vanishes identically, exhibiting thus the behavior of a non-computer state. Although this fact is derived as a smooth limit from the non-extremal RN black hole, it can be directly seen from the Penrose diagram of the extremal case, in which symmetries protect the action in the WdW patch to suffer any evolution (see Figure \ref{Extremal}). From the field theory point of view however, it is not surprising that complexity remains constant. Indeed, extremal black holes have zero temperature, meaning that every property is expected to be static in such states. Nonetheless, extremal black holes provide the first non-trivial state with vanishing computation rate, and might constitute a very  relevant example in the elucidation  of holographic quantum complexity as a microscopic quantity in the CFT side. 

\section{A cold non-computer. Hyperbolic black holes}
\label{sectioncold}

The metric for AdS$_{d+1}$ black holes is given by

\begin{equation}
ds^2 = -f(r) dt^2 + \dfrac{1}{f(r)}dr^2 + r^2\, d\Sigma_{d-1}^2,
\end{equation}
where $d\Sigma_{d-1}$ stands for the spatial $(d-1)$-dimensional boundary metric and we measure length in units of the AdS curvature radius $\ell =1$. The  standard solution is given by

\begin{equation}
f(r) = k+r^2-\dfrac{\mu}{r^{d-2}},
\end{equation}
with $k=0,1,-1$ respectively for flat, spherical and hyperbolic boundary metrics. Thermodynamic properties for these systems can be calculated straightforwardly

\begin{equation}
\label{physicalquantities}
T= \dfrac{dr_h^2+k(d-2)}{4\pi r_h }, \hspace{1.5cm} S = \dfrac{V}{4G} \,r_h^{d-1}, \hspace{1.5cm} M = \dfrac{V (d-1)}{16 \pi G} \mu,
\end{equation}
with $V$ the volume factor in $d-1$ dimensions and $r_h$ the radial location of the horizon. As shown in \cite{BrownSusskindAction}, the late-time complexity growth for these class of solutions yields a universal result for any spacetime dimension, given by the simple formula

\begin{equation}
\label{totalrate}
\dfrac{dI}{dt} = 32\pi G M.
\end{equation}

\begin{figure}[h]
\begin{center}
\includegraphics[height=7cm]{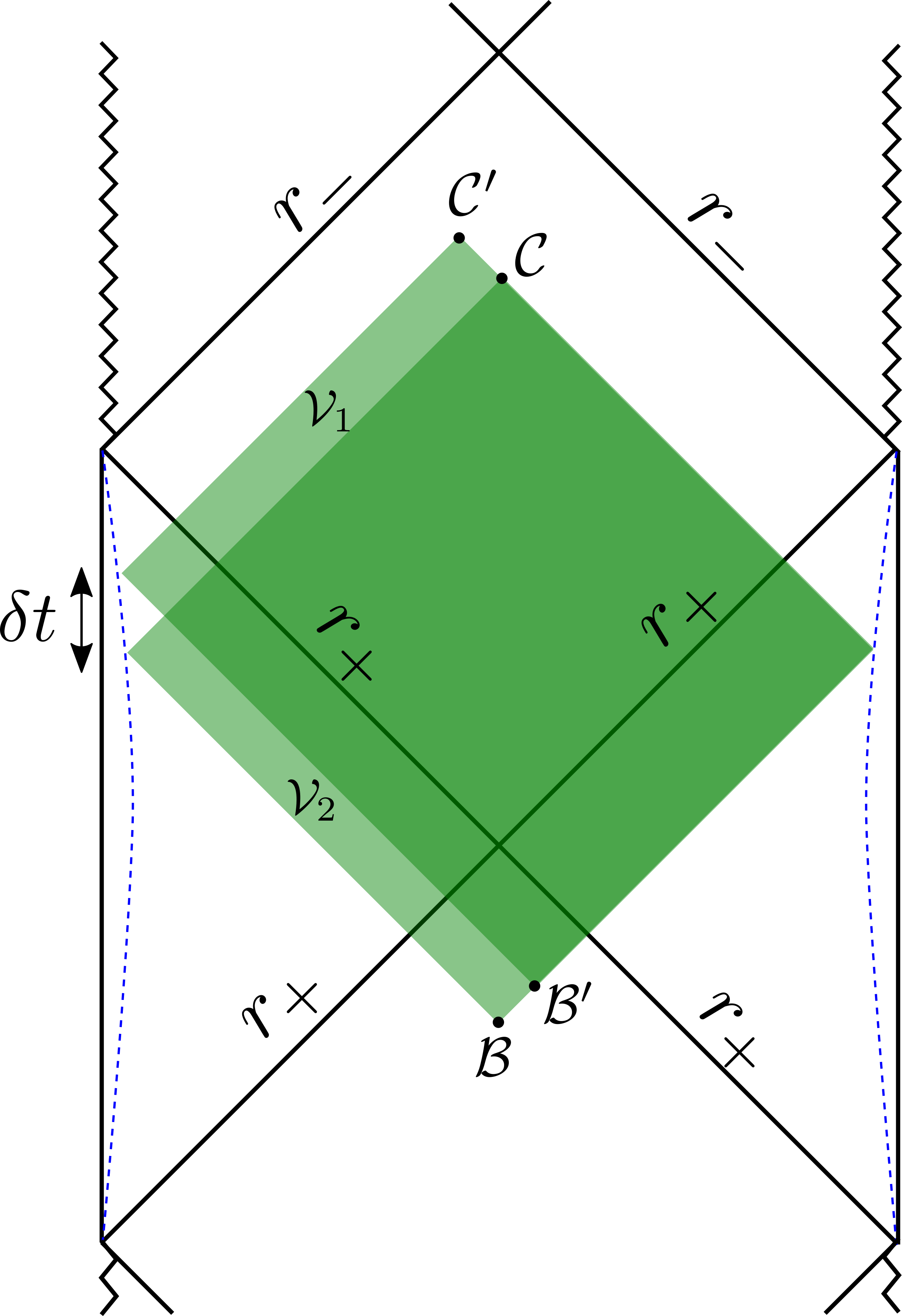}
\end{center}
\caption{WdW patch for a near-extremal (cold) hyperbolic black hole. }
\label{HyperbolicPenrose}
\end{figure}

For the case of hyperbolic black holes (cf. \cite{roberto} for a description), the formula above still holds provided the mass is positive as measured respect to the empty AdS vacuum solution $\mu=0$. Careful analysis of the thermodynamics, however, shows that finite temperature states exist with negative mass respect to this vacuum for the parametric region

\begin{equation}
-\dfrac{2 (d-2)^{\frac{d}{2}-1}}{d^{\frac{d}{2}} } \leq \mu  <0,
\end{equation}
enjoying a conformal diagram whose topology resembles that of Reissner-N\"ordstrom black holes (see Figure \ref{HyperbolicPenrose}). The complexity of these states, characterized by a temperature $0<T<1/2\pi$, is not given by \eqref{totalrate} but requires a separate calculation. Indeed, the computation rate of near-extremal hyperbolic black holes only gets contributions from the joint and bulk pieces of the action, i.e.
the Einstein-Hilbert term

\begin{equation}
\label{coldbulk}
I_{\CV_1}-I_{\CV_2} = -2V(r_\CB^d-r_\CC^d) \delta t,
\end{equation}
and the contribution of the joints

\begin{eqnarray}
\label{coldjoints}
 I_{\CB \CB'} +  I_{\CC \CC'}= \delta t V \left[ 2r^{d}-(2-d)\mu + (d-1)r^{d-2}f(r)\log\left(\dfrac{f(r)}{c \bar c} \right) \right]^{r_{\CB}}_{r_{\CC}}.
\end{eqnarray}
where $c, {\bar c}$ are the conventional normalization parameters of null vectors in the WdW patch (cf. \cite{Poisson}). Adding up \eqref{coldbulk} and \eqref{coldjoints} we get that the total rate is given by

\begin{equation}
\label{coldi}
\dfrac{dI}{dt}=V(d-1) \left[r^{d-2} f(r)\log\left(\dfrac{f(r)}{c \bar c} \right) \right]^{r_{\CB}}_{r_{\CC}}
\;,\end{equation}
which indeed vanishes in the late time limit \footnote{As we see, whether or not this cancellation also holds at early times depends on our choice of normalization $ c, \bar c $ for the null vectors. }, i.e. as $r_{\CB, \CC}  \rightarrow r_{\pm}$. As we see, the above cancellation holds independently of the temperature of the black hole, as long as it lies within the near-extremal regime $0\leq T \leq 1/2\pi$. Thereby, cold hyperbolic black holes provide an example of an ensemble of states enjoying a non-computer behavior. This contrasts with the result obtained within the CV proposal \citep{BarbonMartinHyperbolic} in which the linear growth behavior held also in this regime. 

Despite the finite-temperature nature of these solutions, it must be said  that such states are unlikely to be  stable,  but rather should be interpreted as highly degenerate unstable or perhaps metastable systems. Evidence in this direction comes from the embedding of these solutions into fully fledged string theory systems, such as stacks of type IIB D3-branes, yielding a canonical example of AdS$_5 \times {\bf S}^5$ duality with maximally supersymmetric Yang--Mills theory on an spatial hyperboloid ${\bf H}^{3}$. A marginally tachyonic scalar, saturating the AdS$_5$  BF bound with $m^2 = -4$ and corresponding to BPS-protected scalar mass operators with $\Delta =2$, will have zero modes that actually violate the BF bound in the emerging AdS$_{1+1} \times {\bf H}^3 \times {\bf S}^5$ 
geometry of the near-horizon region of cold hyperbolic black holes.\footnote{We thank B. Freivogel for a discussion on this issue.} In this case the AdS$_{1+1}$ radius of curvature is $\ell' = 1/2$ and the corresponding BF bound $m'^{\,2} \geq -1/4\ell'^{\,2} = -1$. Such systems are therefore expected to undergo tachyonic instabilities. Even if the perturbative instabilities are somehow checked out,  \citep{BarbonTopBH} shows that the cold branch of hyperbolic black holes is likely unstable to non-perturbative D3-brane fragmentation processes.  

It is interesting to notice that the difficulties associated to the emergence of an approximate AdS$_{1+1}$ geometry in the near-horizon region are also responsible for the mismatch between the CA ansatz and the CV ansatz observed for these solutions (cf. \cite{BarbonMartinHyperbolic,MyersFormation}).

\section{A hot non-computer. Large $d$ black holes}
\label{sectionhot}

Away from the late time approximation, higher dimensional black holes are known to exhibit a delay in their computation rate for any $d\geq3$ \cite{BrownSusskindAction}. This phenomenon arises as a consequence of an extra symmetry of the WdW patch at early times which postpones the start of the complexity growth to a later time $t_{\MC}$. In particular, as the spacetime dimension gets bigger, the past and future singularities bow into the Penrose diagram, effectively splitting the time symmetry in two separate left and right time-shift symmetries,  $t_{L,R} \rightarrow  t_{L,R} + c_{L,R}$, for every WdW patch touching both past and future singularities. As the past boundaries leave the singularity, this symmetry breaks down to the smaller boost symmetry $c_L + c_R =0$ and the black hole starts computing. In the following we will calculate the value of these delays and explore its behaviour respect to the spacetime dimension.

\subsection{Computation delays for $d \geq 3$ black holes}
\label{sectiondelays}

We begin by recalling the form of the metric for spherical AdS$_{d+1}$  black holes
\begin{equation}
\label{otram}
ds^2  = -f(r) dt^2 + \dfrac{1}{f(r)}dr^2 + r^2\, d\Omega_{d-1}^2,
\end{equation}
where $d\Omega_{d-1}^2$ is the volume form of the unit ${\bf S}^{d-1}$ sphere with volume
\begin{equation}
\label{sphevol}
V_\Omega = {2 \pi^{d\over 2} \over \Gamma(d/2)}\;,
\end{equation}
and the warping function is given by
\begin{equation}
\label{warp}
f(r) = 1 + {r^2 \over \ell^2 } - \left({r_h\over r}\right)^{d-2}\, \left(1+ {r_h^2 \over \ell^2} \right)\;,
\end{equation}
after we have restored the  dependence on $\ell$, the curvature radius of AdS. 
The basic thermodynamic quantities are given by 
\begin{equation}
\label{physicalquantitie}
T= \dfrac{d\,r_h^2+(d-2) \ell^2 }{4\pi r_h \, \ell^2}, \hspace{1.3cm} S = \dfrac{V_\Omega}{4G} \,r_h^{d-1}, \hspace{1.3cm} M = \dfrac{V_\Omega (d-1)}{16 \pi G} \left(r_h^{d-2} + {r_h^d \over \ell^2}\right)\;.
\end{equation}

In order to study the delay, it is necessary to construct the Kruskal extension for general dimensions. 
The first step will be to define the tortoise coordinate, given by

\begin{equation}
\label{deftortoise}
r_*(r) = \int\limits^{r}_0 \dfrac{dr}{f(r)} +C,
\end{equation}
where the constant $C$ is chosen so that the coordinate is real in the exterior region. Analytic expressions for this integral cannot be found in general. For our purposes however, it will suffice to find the asymptotic limit $r_*(\infty)$, whose value will be crucial in the construction of the conformal diagram. In terms of this coordinate, the Kruskal extension is defined in the lightcone coordinates as follows

\begin{eqnarray}
uv &=& -e^{4 \pi T {r}_*(r)}, \\
\label{uoverv}
\dfrac{u}{v}&=&-e^{4 \pi T t}.
\end{eqnarray}
With this choice, the singularity will be located at $uv=1$ for any dimension, whereas the AdS boundary is located at

\begin{equation}
uv = e^{4 \pi T {r}_*(\infty)}.
\end{equation}
The value of ${r}_*(\infty)$ will in general depend both on the dimension and physical parameters of the solution, with qualitatively different behaviors depending on the relative size of the black hole respect to the curvature radius.

\subsubsection{Large AdS black holes}

In the large black hole limit\footnote{For flat ($k=0$) AdS Black holes, this condition is not needed and the result holds for any hierarchy of $r_h$ and $\ell$.} ($r_h \gg \ell$) we might approximate

\begin{equation}
f(r) = 1 + \dfrac{r^2}{\ell^2} - \left( \dfrac{r_h}{r} \right)^{d-2} \left(1+ \dfrac{r_h^2}{\ell^2} \right) \simeq \dfrac{r^2}{\ell^2} - \dfrac{r_h^d}{\ell^2 r^{d-2}},
\end{equation}
and we can calculate the integral in \eqref{deftortoise} analytically 

\begin{equation}
\int\limits^{r}_{0} \dfrac{dr}{f(r)} = \frac{\ell^2}{r} \left[_2F_1\left(1,-\frac{1}{d};1-\frac{1}{d};\left(\frac{r}{r_h } \right)^d\right)-1\right]\,,
\end{equation}
which forces us to choose $C= -i \frac{\pi}{d}$. Using the asymptotic expansions for the hypergeometric function at large $r/r_h$ we get (cf. \cite{Kim, Myerstdep})

\begin{equation}
\lim\limits_{r \rightarrow \infty} r_{*}(r) = \dfrac{1}{4 T} \cot \dfrac{\pi}{d}.
\end{equation}
As we see, the value of $uv$ at the boundary depends only on the dimension for large black holes

\begin{equation}
uv = e^{\alpha(d)},
\end{equation}
with $\alpha(d)=\pi \cot \frac{\pi}{d}$, an approximately linear function of $d$. This means that as $d$ grows, the corresponding hyperbola in the Kruskal diagram is further apart from the origin. In order to construct now the Penrose diagram, we might choose to flatten one of the two pairs of hyperbolas. If we choose (as usual) to flaten the AdS asymptotic boundary, we may perform the change of coordinates

\begin{eqnarray}
\label{Penrosecoordinates}
v &=& e^{\frac{\alpha(d) }{2}} \tan \dfrac{\tilde{v}}{2}, \\
u &=& e^{\frac{\alpha(d) }{2}} \tan \dfrac{\tilde{u}}{2},
\end{eqnarray}
in which the full spacetime is now compactified in a finite region. Undoing the lightcone coordinates

\begin{eqnarray}
\tilde{u} &=& \tau + \rho, \\
\tilde{v} &=& \tau-\rho,
\end{eqnarray}
it is easy to see that the AdS boundary at $uv=e^{\alpha(d)}$ is now given by the straight lines $\rho= \pm \frac{\pi}{2}$. The singularity, on the other hand, becomes bowed in \footnote{Had we chosen to flatten the singularity in the Penrose diagram, the result would have been that the AdS boundary becomes bowed out. One might wonder if there exist a suitable conformal transformation that could flatten out both boundaries at the same time. Symmetries guarantee that this is not possible in this case \cite{Hubeny}.} with a form given implicitly by

\begin{equation}
\tan\left(\dfrac{\tau+\rho}{2} \right) \tan\left(\dfrac{\tau-\rho}{2} \right) = e^{-\alpha(d)}.
\end{equation}
Assuming a symmetric evolution for the action growth (i.e. the WdW patch starts at the same asymptotic time in both sides $t_L=t_R$ ), it is possible to calculate the time at which the 'south tip' of the WdW patch leaves the singularity for the first time. This will correspond to the time at which the black hole starts computing. Finding the intersection of the past singularity with $\rho=0$ and inverting back to the asymptotic time $t$ we get that the delay is given by the simple expression

\begin{equation}
\label{bigdelay}
t_\mathcal{C} = \dfrac{\alpha(d)}{4 \pi T} \simeq \dfrac{d}{4 \pi T} + \mathcal{O}(1/d).
\end{equation}
As we could have intuitively expected, the computation delay increases as the singularity bows further into the diagram for higher and higher dimensions. As the spacetime dimension changes, however, many physical properties of the black hole might become trivial unless the scales in the problem are made $d$-dependent. The latter interpretation, thereby, can depend on such possible scalings. In section \ref{sectionscalings}, we will discuss such scalings and their implications in the study of complexity for large $d$ black holes.

\begin{figure}[h]
\begin{center}
\includegraphics[height=7cm]{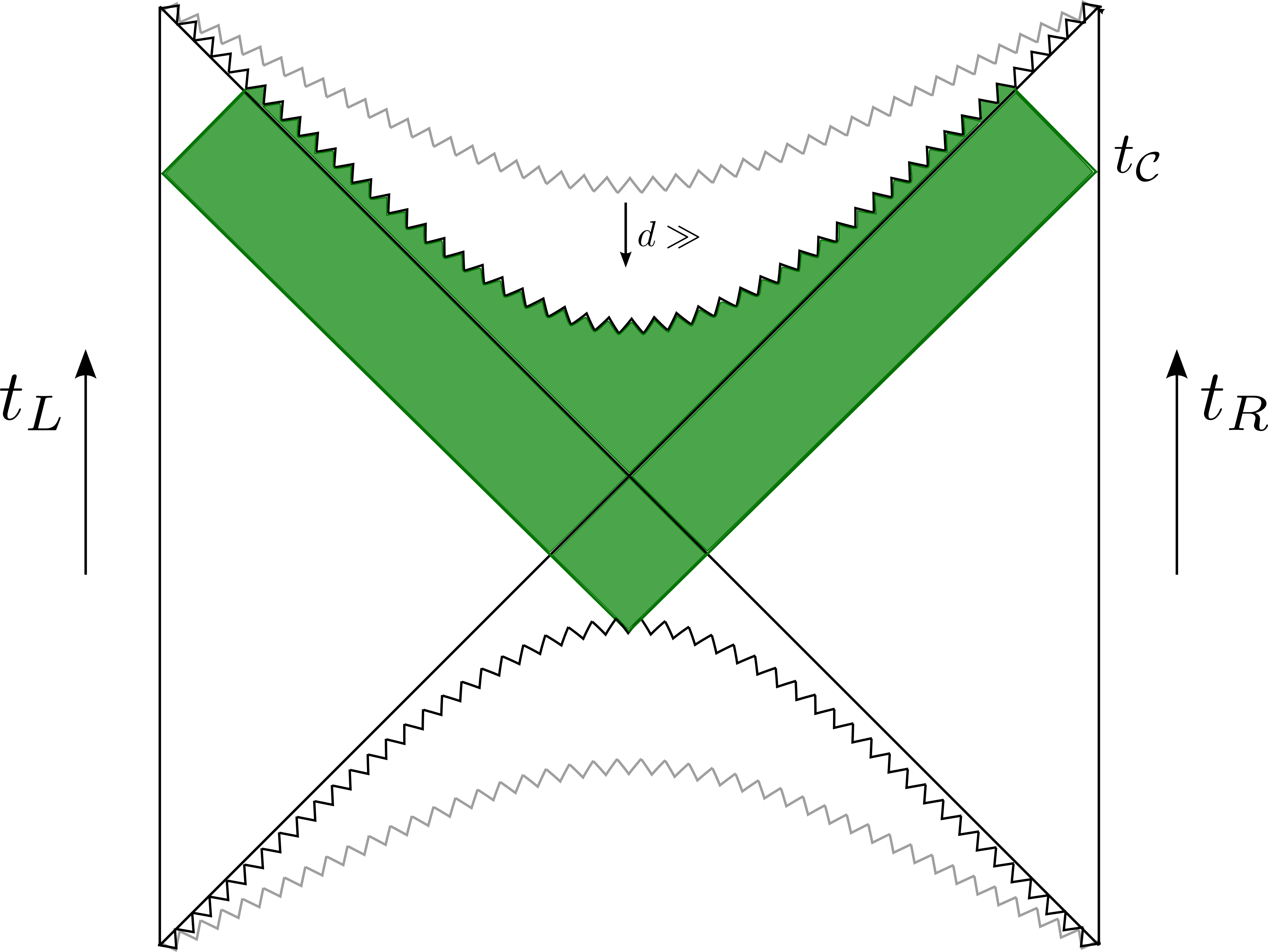}
\end{center}
\caption{Conformal diagram for higher dimensional black holes and WdW patch at the moment of computation starting. }
\label{penroselargeD}
\end{figure}

\subsubsection{Small AdS black holes. Schwarzschild black holes.}

The shape of the large-$d$ conformal diagram  shows some significant differences for the case of small AdS black holes. In particular, the singularity does not bow arbitrarily further in as the dimension grows, and the tortoise coordinate asymptotic value saturates at the same value independently of $d$. In particular, for $r_h \ll \ell$ we may approximate $f(r) \simeq 1+\frac{r^2}{\ell^2},$ and using the definition in \eqref{deftortoise} we get

\begin{equation}
r_*(r) = \ell \arctan \frac{r}{\ell},
\end{equation}
whose $r\rightarrow\infty$ limit gives us the corresponding delay

\begin{equation}
\label{smalldelay}
t_\mathcal{C} \simeq \dfrac{\pi}{2} \ell.
\end{equation}
Equivalently, asymptotically flat Schwarzschild black holes in a box give a similar solution, i.e. a computational delay that is only controlled by the size of the box. Indeed, for $f(r) = 1- (r_h/r)^{d-2}$ we get

\begin{equation}
r_*(r) = r \, _2F_1\left(1,\frac{1}{2-d};1+\frac{1}{2-d};\left(\frac{r_h}{r}\right)^{d-2} \right)\;.
\end{equation}
If we regard the WdW patch as anchored at the walls of the box, we must evaluate the tortoise coordinate at the location of the box in order to find the corresponding delay. For a well-contained black hole, $L \gg r_h$, we have
the asymptotic behavior $r_* (L) \sim L$ and we obtain 
$$
t_\mathcal{C} = r_* (L) \simeq L\;.
$$
We see that well-contained black holes have computation delays controlled by the  size of box rather than the black hole itself. In other words, the non-computing feature is a property of the combined system, including both the black hole and its `container'. 

\begin{figure}[h]
\begin{center}
\includegraphics[height=6cm]{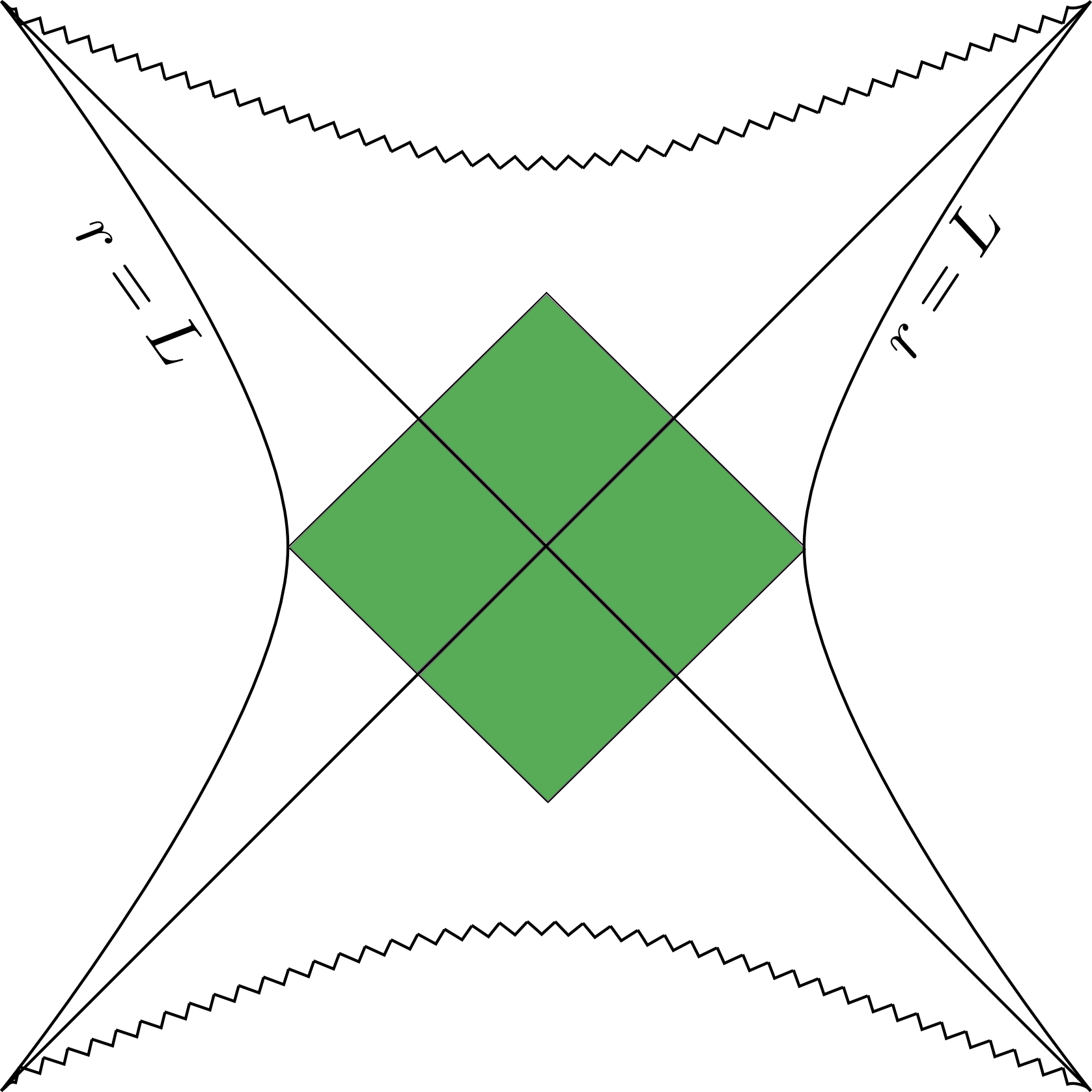}
\end{center}
\caption{If a box is too small ($L< L_c$) the non-computing features can disappear.}
\label{fig:smallbox}
\end{figure}

It is then interesting to ask what happens when we shrink the box down to the size of the black hole. For AdS black holes, there is
a smooth transition from small to large black holes. At the transition region we have $T \sim 1/\ell$, so that the small black-hole behavior 
(\ref{smalldelay}) smoothly morphs into the large black-hole behavior (\ref{bigdelay}). On the other hand, for asymptotically flat  Schwarzschild black holes with a WdW patch anchored at the location of walls, there is always a limiting value of the box size, $L_c$,  below which the combined system of black hole and box cease to present non-computer behavior, since the WdW patch eventually becomes too small to simultaneously touch both past and future singularities (cf. Figure \ref{fig:smallbox}). The smallest WdW patch with a non-computer delay is anchored at a zero of the tortoise coordinate, i.e.
$$
r_* (L_c) = 0
$$

\section{Holographic non-computers in the large $d$ limit}
\label{sectionscalings}

In the previous section we have seen that large AdS black holes feature a computational delay which becomes parametrically large at large dimensions. This suggests the analysis of holographic complexity in the $1/d$ expansion of general relativity \cite{EmparanlargeD, Minwalla}. These large-$d$ approximations are a kind of mean-field expansion which reveal interesting structure in many classical gravitational phenomena.  A non-trivial question is whether there exist a set of large-$d$ scalings which preserve the standard `phenomenology' of complexity, namely the existence of a linear growth and a large-complexity saturation at very long times (cf. \cite{Susskindtypical}). 

\begin{figure}[h]
\begin{center}
\includegraphics[height=4.5cm]{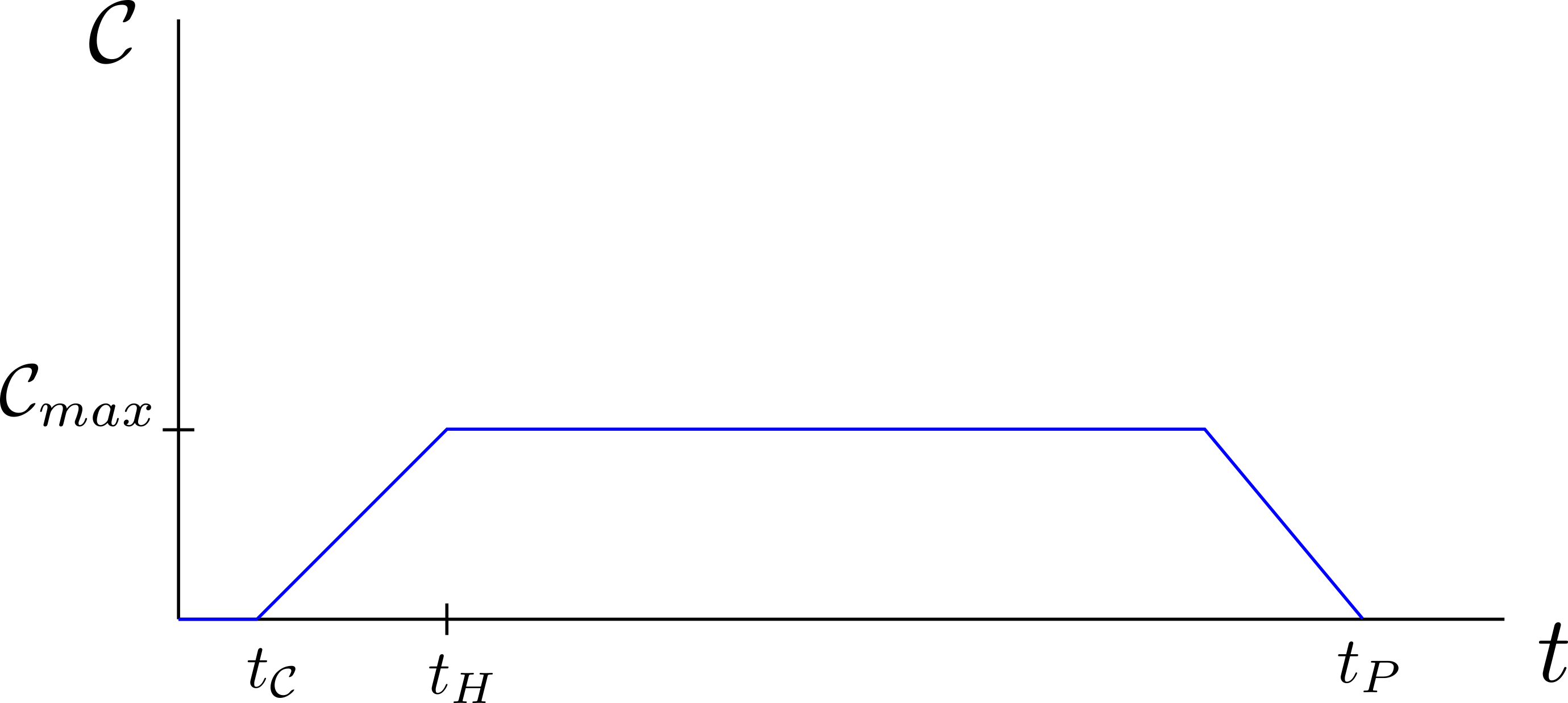}
\end{center}
\caption{Scheme for complexity pattern of a finite $d$ black hole.}
\label{plateau}
\end{figure}

The holographic prescription captures the growth of complexity at a rate of order $ST \sim M$, up until we reach complexities of order
\begin{equation}
C_{\rm max} \sim \log(1/\epsilon) \,e^S\;,
\end{equation}
where $\epsilon$ is a coarse-graining parameter in Hilbert space, controlling the degree of approximation we require to `stop the computation'. It is unclear to what extent $\epsilon$ could have an interpretation in the bulk geometry. Assuming $\log (1/\epsilon)$ of order unity, the time of complexity saturation is thus of the order of the Heisenberg time of the system, $t_H \sim T^{-1} e^{S}$ up to subleading terms in the exponent. Over periods of the order of the quantum Poincar\'e recurrence time, $t_P \sim T^{-1} \exp\left(e^S \log(1/\epsilon) \right)$, one expects the system to undo its evolution and decrease its complexity.   A caricature of this behavior is shown in figure \ref{plateau}. Notice that any large-$d$ scaling preserving the plateau shape must keep finite both the mass and the entropy of the black hole.  

Making this choice however implies that we should not forget about the Hawking process, which actually becomes rather violent in the large $d$ limit. Indeed, typical frequencies for Hawking quanta scale as $\omega \sim d^2/r_h$ , yielding a radiation power that grows factorially in $d$ \cite{Hod} and implying thus an almost immediate evaporation in the large $d$ limit. Of course, a suitable scaling of quantities could be made in order to keep the evaporation time finite, but would limit our possibilities to do so with other quantities of interest. Instead, we will require our `computers' to remain   in thermodynamic equilibrium for exponentially long times, such as large AdS black holes or Schwarzschild black holes inside a suitable box,  in order to avoid the evaporation process. 

In the following sections we show that the requirement of thermodynamic stability is actually non-trivial for small Schwarzschild black holes. On the other hand, no  obstructions are found for large AdS black holes, which are always thermodynamically stable. 

\subsection{$1/d$ scaling for large AdS black holes}

In the large AdS black hole regime, $r_h \gg \ell$, the relevant thermodynamic  quantities behave as follows

\begin{equation}
T= \dfrac{dr_h}{4\pi \ell^2}, \hspace{1.5cm} S = \dfrac{V_\Omega}{4G} r_h^{d-1}, \hspace{1.5cm} M = \dfrac{V_\Omega (d-1)}{16 \pi G \ell ^2} r_h^d,
\end{equation}
satisfying the relation

\begin{equation}
\label{smarr1}
TS = \dfrac{d}{d-1} M,
\end{equation}
which stabilizes in the large $d$ limit. Since keeping the parameters $M$ and $S$ is fundamental to maintain the plateau-shape of the  complexity function, \eqref{smarr1} forces us to fix the temperature as well for the picture to be consistent. To do so, we may rewrite the temperature as

\begin{equation}
T= \dfrac{d}{4\pi} \left( \dfrac{r_h}{\ell} \right) \dfrac{1}{\ell}.
\end{equation}

Now, making sure that we stay in the large black hole regime implies that the ratio $r_h/\ell$ is always above unity. Thus, we might choose a general set of scalings

\begin{equation}
\label{fded}
\dfrac{r_h}{\ell} = f(d),
\end{equation}
with $f(d)$ either a constant larger than unity or a growing function of $d$. Once this function is chosen, we must rescale the AdS radius in such a way that the temperature remains finite, i.e.

\begin{equation}
\ell \,T \sim  d\, f(d) ,
\end{equation}
To make the entropy finite we can now exploit our freedom to rescale the Planck length $G= (\ell_P)^{d-1}$ as 

\begin{equation}
\left( \dfrac{\ell}{\ell_P} \right)^{d-1} \sim \dfrac{S}{f(d)^{d-1} \,V_\Omega},
\end{equation}
with fixed $S$. 
In order to ensure consistency of the geometrical  description, $\ell_P \ll \ell$, we must limit the growth of $f(d)$ to remain below $O(\sqrt{d})$, since then the strong vanishing of the unit volume $V_\Omega \rightarrow \left({1/\sqrt{d}}\right)^d$ is enough to maintain $\ell_P$ as the hierarchically smaller length scale in the problem. 
 
Once we stabilize the scalings of $S$ and $T$, the mass $M$ is kept stable by the Smarr relation \eqref{smarr1}, thus keeping  the qualitative shape of the plateau as $d$ becomes large. Looking now at the computational delay, we can see that the finiteness of the temperature ensures that $t_{\mathcal{C}}$ blows up as $d$ becomes large, meaning that the complexity plateau becomes postponed away in the future

\begin{equation}
t_{\mathcal{C}} \sim \dfrac{d}{T}.
\end{equation}

At leading order in the $1/d$ expansion, we have thus  a parametric example of a holographic non-computer, i.e. a finite temperature state for which complexity seems to remain always at a constant value.

\begin{figure}[h]
\begin{center}
\includegraphics[height=4cm]{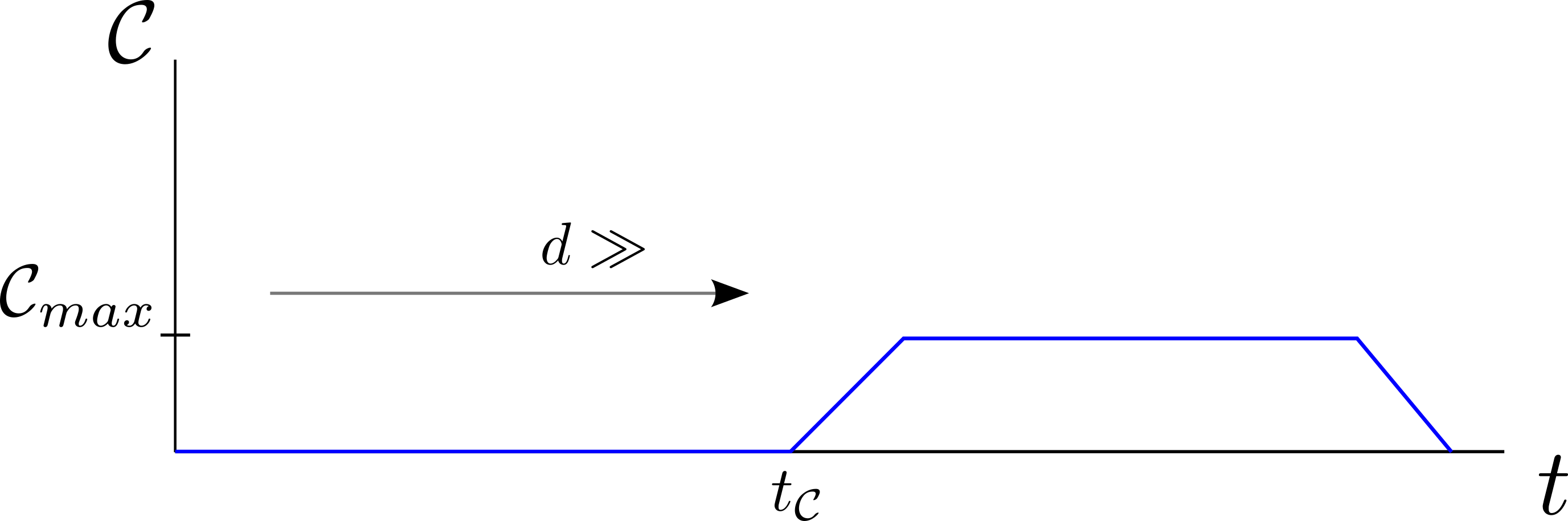}
\end{center}
\caption{Plateau shifting of large $d$ AdS black holes.}
\label{plateau2}
\end{figure}

\subsection{Infinite-volume scaling}

It is interesting that the successful large-$d$ scaling of the complexity plateau involves a large-$d$ scaling of the  AdS `containment box' .   Since the radius of AdS becomes a physical box size in the CFT dual, it is interesting to reformulate the problem in terms of a `complexity density' which becomes stable in the infinite-volume limit of the CFT.  To this end we consider   black-brane solutions dual to thermal states on flat space.     Now we are free from any restrictions regarding the $r_h/\ell$ ratio, and  this additional freedom allows us to preserve the complexity plateau without scaling with $d$ every physical scale in the problem. 

The temperature formula 
\begin{equation}
T = {d \,r_h \over 4\pi \ell^2}
\end{equation}
remains the same as before. However, the horizon entropy  {\it density} is now given by
\begin{equation}
{S \over V}= {1 \over 4G} \, \left({r_h \over \ell}\right)^{d-1} = N_* \, \left({4\pi T \over d}\right)^{d-1} \;,
\end{equation}
where we have denoted $N_* = \ell^{d-1} /4G$ the effective number of `species' in the CFT (proportional to the central charge). 

At fixed $T$, the power-like behavior proportional to $T^{d-1}$ implies that any notion of entropy which remains stable in the large-$d$
limit must factor out this term. A natural way of achieving this is to focus on the entropy per thermal cell, namely
\begin{equation}
S_{\rm cell} \equiv {S \over VT^{d-1}}\;, 
\end{equation}
and a similar definition for the thermal-cell energy:
\begin{equation}
M_{\rm cell} \equiv {M \over VT^{d-1}}\;.
\end{equation}
Scaling now $N_* \rightarrow \infty $ according to 
\begin{equation}
N_* = S_{\rm cell}   \left( \dfrac{d}{4 \pi}\right)^{d-1},
\end{equation}
as $d\rightarrow \infty$ with  fixed $S_{\rm cell}$, we make stable the `thermal cell complexity' given by 
\begin{equation}
C_{\rm cell} \equiv {C \over V T^{d-1}}\;.
\end{equation}
Then we find that $C_{\rm cell}$ should reproduce a plateau shape with parameters $S_{\rm cell}, T$ and $M_{\rm cell}$. 
As before, the delay time remains given by $t_{\cal C} \sim d/T$, which diverges linearly in the large $d$ limit.

\subsection{$1/d$ scaling for small  AdS black holes}

If we now consider small black holes in AdS, $r_h \ll \ell$, the thermodynamic quantities will behave as those of the the usual Schwarzschild black holes

\begin{equation}
T= \dfrac{d-2}{4\pi r_h}, \hspace{1.5cm} S = \dfrac{V_\Omega}{4G} r_h^{d-1}, \hspace{1.5cm}  M = \dfrac{V_\Omega (d-1)}{16 \pi G } r_h^{d-2}.
\end{equation}
And again, we can find a simple expression relating the three of them which stabilizes in the large $d$ limit
\begin{equation}
\label{smarr}
TS = \dfrac{d-2}{d-1} M.
\end{equation}
Keeping now a finite temperature requires that we scale up the horizon radius as 
\begin{equation}
r_h  \,T \sim  d, 
\end{equation}
whereas the entropy $S$  is fixed if we scale the Planck length as 
\begin{equation}
\label{planckscaling}
\left( \dfrac{r_h }{\ell_P} \right)^{d-1} \sim \dfrac{{S}}{V_\Omega}. 
\end{equation}

Up to this point, the AdS radius did not make an appearance. However, it will be the relevant scale for the computation delay \eqref{smalldelay} and it is constrained by the requirement that the black hole actually `fits the box', i.e. $r_h < \ell$. In general we can allow 
\begin{equation}
\left( \dfrac{r_h}{\ell} \right) = g(d),
\end{equation}
with $g(d)$ either a small constant or a decreasing function of $d$. Feeding these scalings into \eqref{smalldelay} we get a computation delay
\begin{equation}
t_\mathcal{C} \sim \dfrac{d}{g(d)} T^{-1},
\end{equation}
which again diverges in the large $d$ limit for any of the allowed behaviours of $g(d)$. The case of Schwarzschild black holes
well-contained in a flat box follows along similar lines, with the size of the box playing the role of the AdS radius, $\ell$. 

Our analysis shows that a blow-up of the `containment box' is essential to manufacture a large-$d$ Schwarzschild non-computer. Since
thermodynamic equilibrium of ordinary black holes in finite boxes requires certain ratios between the relative sizes of the black hole and the box, we must check the compatibility of stability  with the required  large-$d$ scaling. 

 \subsubsection{Stability analysis at large $d$}

 Having isolated large-$d$ scalings with parametric computational delay for both large and small black holes, we come now to the discussion of their thermodynamical stability. Since the discussion of action-complexity is formally tied to the two-sided eternal black hole geometries, we shall focus mostly on the canonical ensemble at fixed temperature, which is the effective one-sided description of the associated thermofield-double states. 
 
 The canonical thermodynamics for AdS black holes is well known (cf. \cite{page}). Large and small black holes form a continuous family of solutions labeled by the horizon radius $r_h$. For $r_h \ll \ell$ all small black holes have negative specific heat and their thermodynamics is  locally unstable. The associated temperature is large, and the dominant phase in this regime is a large AdS black hole with $r_h \gg \ell$ and positive specific heat. There is a critical temperature, the so-called Hawking--Page (HP) temperature, $T_{\rm HP} = (d-1)/2\pi \ell$,  below which the large AdS black hole has larger free energy than a gas of gravitons in AdS. Below the HP temperature there is a narrow window down to $T_l = \sqrt{d(d-2)} /2\pi \ell$ in which black holes are locally stable but globally unstable. In this narrow window the size of the black holes is of order $\ell$ and all of them have computational delays of order $\ell$. 
 
Locally stable but globally unstable  entangled black holes should behave as ordinary holographic computers for large periods of time, exponential
in $1/G$, where $G$ is Newton's constant, after which they are likely to fluctuate into a state of two entangled boxes filled with radiation, with a complexity of order $G^0$. It would be very interesting to study how this time scale compares to the Heisenberg time scale, controlling the saturation of complexity. At any rate, black holes whose thermodynamic state is both locally and globally stable, i.e. those with $r_h >\ell$,  are guaranteed to last beyond the saturation plateau and furnish the pattern of large-$d$ computational delay indicated in the previous section. 

The situation is different for asymptotically-flat Schwarzschild black holes contained inside entangled cages of size $L$. If each black hole is much
smaller than its cage, it is guaranteed to be locally unstable, so that it will decay very fast into a graviton-gas state (cf. \cite{gibbons-perry}). In the present interpretation, we say that the thermofield double state will look like an entangled pair of boxes full of radiation for almost all the time. Such states should have growing complexity of order $G^0$. On the other hand, for black holes which almost touch the cage, there are windows of local and global stability for growing complexity of order $1/G$. Following   \cite{York}, we can
determine these regimes by evaluating the Euclidean action of the black hole solution with two boundary conditions: the temperature is physically fixed at the walls of the box for both the black hole and the graviton gas states, and of course the metric is smooth at the horizon. 

Writing  the Euclidean black hole metric as 
$$
ds^2_{\rm bh} = \left(1-\left(\frac{r_h}{r} \right)^{d-2} \right) \,d\tau'^{\,2} + \dfrac{dr^2}{\left(1-\left( \frac{r_h}{r} \right)^{d-2} \right)}   + r^2 d\Omega_{d-1}^2\;,
$$
with $\tau' \equiv \tau' + \beta'$, we require that the ${\bf S}^1$ parametrized by $\tau'$ be smoothly contractible, which fixes 
$$
\beta' = {4\pi r_h \over d-2}\;.
$$
On the other hand, the physical temperature is measured as the inverse proper length of the ${\bf S}^1$ at the walls of the box, i.e.
\begin{equation}
\label{smo}
\beta = {1\over T} = \beta' \sqrt{ 1- (r_h /L)^{d-2}}\;. 
\end{equation}
The vacuum metric which is used for normalization is given by
$$
ds^2_{\rm vac} = d\tau^2 + dr^2 + r^2 d\Omega_{d-1}^2\;,
$$
with $\tau \equiv \tau + \beta$. The canonical free energy is computed in the saddle-point approximation by substracting the corresponding  Euclidean actions. Ricci flatness of both solutions implies that only the YGH term contributes in both cases:
$$
-\log Z(\beta) \approx -{1\over 8\pi G} \int_{\partial {X_{\rm bh}}} K + {1\over 8\pi G} \int_{\partial {X_{\rm vac}}} K = \beta M_{\rm eff} - S\;,
$$
 where $S$ is the entropy of the black hole and $M_{\rm eff}$ is the quasilocal Brown-York mass (cf. \cite{brown-york}) given by
 \begin{equation}
 \label{meff}
 M_{\rm eff} = 2L^{d-2} {(d-1) \Omega_{d-1} \over 16\pi G} \left(1-\sqrt{1-(r_h /L)^{d-2}}\right)\;.
 \end{equation}
 Notice that this effective mass approaches the standard ADM mass of the black hole as we push the cage to infinity, $L\rightarrow \infty$. The form of  $M_{\rm eff}$ is completely fixed by the Bekenstein--Hawking formula
 $$
 S= {\Omega_{d-1} \over 4G} r_h^{d-1}\;,
 $$
 together with the smoothness condition (\ref{smo}). To see this, notice that we can rewrite the first law as
 $$
 \beta = {\partial S \over \partial E} = {d S \over d r_h} {\partial r_h \over \partial E}\;,
 $$
 where $E$ is the internal energy. 
 Since we know the functional dependence of both $\beta$ and $S$ on $r_h$, the previous relation is a simple differential equation for
 $E(r_h)$. This equation is easily solved with the condition that $E(r_h =0) =0$ to yield exactly the expression (\ref{meff}): 
 $$
 E(r_h) = M_{\rm eff} (r_h)\;,
 $$
 and the free energy follows then from the standard thermodynamic relation
 $$
 \log Z(\beta) = -\beta E + S\;.
 $$
 At any rate, our expression for $\log Z(\beta)$ as a function of $r_h$ determines a window of local stability for black holes which
 are sufficiently close to the walls of box. In terms of the parameter 
 $$
 x\equiv \left({r_h \over L}\right)^{d-2}\;,
 $$
 locally stable black holes exist inside the cage for $x_l < x < 1$ with 
 $$
 x_l = {2\over d}\;.
 $$
 Globally stable black holes are determined by a negative free energy, which requires  that $x_s < x< 1$ with
 $$
 x_s = 4{d-1 \over d^2}\;.
 $$
 Notice that, as $d\rightarrow \infty$, the stable black holes lie arbitrarily close to the walls of the box. 
 
\begin{figure}[h]
\begin{center}
\includegraphics[height=5cm]{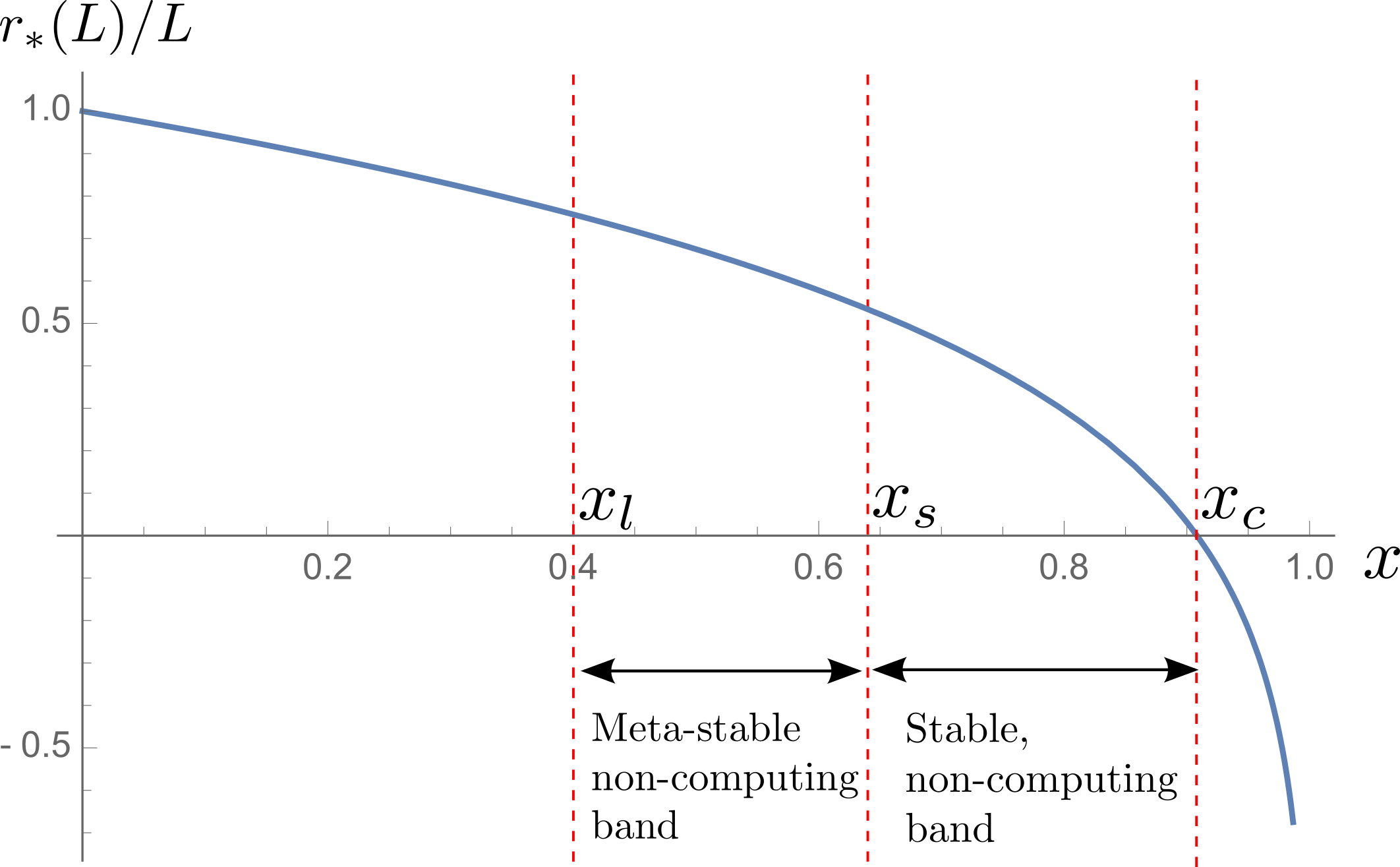}
\end{center}
\caption{Tortoise coordinate at the wall as function of $x=\left({r_h \over L}\right)^{d-2} $. The band compatible with global stability as well as non-computing features lies within $x_c< x < x_s$. }
\label{fig:hyper}
\end{figure}

 These windows of stability combine in a non-trivial fashion with the requirement that they behave as holographic non-computers. As indicated in the previous section, the condition for the black hole to possess a computational delay is that the cage is not too small. In particular, the critical value for non-computing, determined by $r_*(L_c) =0$ must be such that $x_c = (r_h /L_c)^{d-2}$ be {\it smaller} than $x_s$. Only then we can find stable black holes with a non-computing WdW patch. Alternatively, we require that the tortoise coordinate at the wall be {\it positive} for the critically stable black hole at $x=x_s$. We show in Figure \ref{fig:hyper} that this is indeed the case, so that a  band of large-$d$ non-computers exist among the narrowly caged Schawrzschild black holes. 
 
\subsection{Firewalls as natural non-computers?}
\label{sectionfirewalls}

As we learned in section \ref{sectiondelays}, singularities of some black hole solutions become arbitrarily close to the horizon in the large $d$ limit, suggesting the fact that large $d$ black holes could provide a classical model of firewalls \citep{AMPS}. The exotic complexity dynamics of such solutions raises the question of whether firewalls might actually provide a natural candidate of non-computer systems. 

In order to check if large-$d$ black holes are really classical models of firewalls, we must check if the physical `thickness' of the black-hole interior is Planckian.  We can phrase this question  by  calculating the proper free-fall time through the interior geometry,  towards the singularity. For big (flat and spherical) AdS black holes, this is given by

\begin{equation}
\label{bigcrash}
\tau_{sing} = \int\limits^{r_h}_0 \dfrac{dr}{\sqrt{-f(r)}} \simeq \dfrac{\pi}{d} \ell,
\end{equation}
whereas this quantity is controlled by the size of the horizon for small AdS and Schwarzschild black holes

\begin{equation}
\label{smallcrash}
\tau_{sing} \simeq \dfrac{r_h}{d}.
\end{equation}

A classical model for a firewall would presumably correspond to a Planckian infalling time towards the singularity. The particular scalings defined in this paper, which are fixed by the requirement of keeping a qualitative plateau-shape for the complexity growth, imply an effective shrinking of the Planck length, so that the falling time is always large compared to the Planck length in the case of finite-entropy black holes (large or small). For the case of large AdS black holes we have
\begin{equation}
\left({\tau_{sing} \over \ell_P}\right)^{d-1}  \sim {S \over (d\, f(d))^{d-1} V_\Omega} \sim \left({\sqrt{d} \over f(d)}\right)^{d-1},
\end{equation}
which diverges at large $d$, under the condition $f(d) < \sqrt{d}$, which was imposed to ensure that the Planck length is indeed smaller than the AdS radius. In the case of small AdS black holes, a similar estimate yields a scaling proportional to $(\sqrt{d}\,)^{d-1}$, which again diverges as $d\rightarrow \infty$. Hence, we conclude that the large-$d$ `shrinking' of the interior geometry is not felt by an infalling observer as a Planckian wall. 

On the other hand, it is interesting to point out that for flat branes we do not need to scale $\ell$ in order to achieve stable `thermal-cell complexity'. In this case we can actually bring a `firewall' physically close to the horizon while maintaining the shape of the plateau. It would be interesting to study if these considerations have any significance for the  meaning of `firewall' states.

\section{Discussion}

We have studied the phenomenology of Action-Complexity dynamics for a class of black hole solutions presenting constant values for the computational complexity, circumventing the expected growing behavior for thermal quantum systems. In this context, we distinguish three types of systems: zero temperature states, cold degenerate states and hot stable states. 

The first example, corresponding to charged extremal black holes, confirms the expected intuition for zero temperature systems, providing a completely static system in which all physical properties, including complexity, remain constant.
The second case, illustrated by cold hyperbolic black holes, provides the first example of a finite-temperature system with constant complexity, contradicting the general expectations for quantum systems. The origin of this behavior, as well as its discrepancy with the Volume-Complexity proposal, are not well understood. Nevertheless, instabilities appearing in consistent string theory embeddings of these systems could have a decisive impact on the prediction, and its understanding could lead to a clarification of the exotic properties of these degenerate systems.

Finally, we show that a formal application of the large-$d$ expansion of GR to large AdS black holes produces parametric examples of holographic non-computers  with computational delays scaling linearly with $d$. From the gravitational point of view, the origin of this phenomenon can be traced back to the existence of a larger set of   independent symmetries acting on the WdW patches  for $t< t_{\mathcal{C}}$. 

We find that small Schwarzschild black holes are somewhat puzzling. First of all, their computational delay does not appear to be intrinsic, but rather depends on  the infrared regulator, i.e. the containment box. Despite the apparent existence of a parametric delay of ${\cal O}(d)$ in the large-$d$ limit, one ultimately finds this incompatible with the requirement that the black hole be stable  unless we fine tune the walls of the box to approach the horizon as $d\rightarrow \infty$. Otherwise we are left with a trivial realization of the `non-computer' in this case, namely two entangled boxes full of radiation.  

 It is interesting to notice that the complexity-phenomenology discussed in this paper  seems certainly particular to the CA conjecture, and does not appear (at least in an obvious manner) in the older CV proposal.  In this sense, it joins the properties of cold hyperbolic black  holes in the list of identifiable discrepancies between the two proposals, a question which deserves further scrutiny.

A major open problem is the understanding of the various non-computing systems described here in the language of the CFT. On general grounds, we expect the large $d$ limit of gravity to correspond to the mean field theory approximation of QFT. In this context, it might be not so surprising that some fine grained properties of the field theory, such as complexity, are not captured by this approximation, yielding a completely trivial dynamics for the leading order in the $1/d$ expansion. On the other hand, given the scarcity of CFTs in higher dimensions, the very existence of a parametric $1/d$ expansion in the AdS/CFT correspondence is a rather intriguing, albeit remote possibility. 

\subsubsection*{Acknowledgements}

We would like to thank C. Gomez, E. Rabinovici and J. R. Laguna for discussions on various aspects of computational complexity.  This work is partially supported by the Spanish Research Agency (Agencia Estatal de Investigaci\'on) through the grants IFT Centro de Excelencia Severo Ochoa SEV-2016-0597 and FPA2015-65480-P. The work of J.M.G. is funded by Fundaci\'on La Caixa under ``La Caixa-Severo Ochoa'' international predoctoral grant. 


\bibliographystyle{utphys.bst}
\bibliography{refs}{}
\end{document}